\documentclass[aps,prl,preprint,showpacs,superscriptaddress]{revtex4}

\begin{document}

\preprint{KA--TP--06--2003}

\title{\vspace*{2\baselineskip}
       Comment on ``Vacuum Photon Splitting in Lorentz-Violating
       Quantum Electrodynamics''
       \vspace*{1\baselineskip}}

\author{C. Adam}
\email{adam@radon.mat.univie.ac.at}
\affiliation{Wolfgang Pauli Institut, c/o Institut f\"ur Mathematik,\\ 
             Universit\"at Wien, 1090 Wien, Austria}

\author{F.R. Klinkhamer}
\email{frans.klinkhamer@physik.uni-karlsruhe.de}
\affiliation{Institut f\"ur Theoretische Physik,
             Universit\"at Karlsruhe (TH),\\ 76128 Karlsruhe, Germany 
             \vspace*{1\baselineskip}}

\date{August 2, 2003}   


\pacs{12.20.-m, 11.15.Bt, 11.30.Cp, 11.30.Er}

\maketitle

In a recent Letter \cite{KoPi}, Kosteleck\'{y} and Pickering argue 
that there may be observable effects from photon 
triple splitting  in an extended version of quantum electrodynamics with Lorentz 
violation in the fermion sector. The argument is based on 
a supposed analogy with a known physical process.
In this Comment, we point out that the analogy is misleading and 
that, at the order considered, 
the probability of on-shell photon triple splitting is strictly zero.

Kosteleck\'{y} and Pickering \cite{KoPi} study the process of photon
splitting in certain Lorentz- and CPT-violating extensions of 
quantum electrodynamics. For conventional quantum electrodynamics, the
photon splitting amplitude is zero to all orders of perturbation theory, 
but in a Lorentz-violating extension of the theory the situation may be
different. Concretely, the authors study a model where the photon
sector is conventional and the Lorentz violation is due to
additional terms in the fermion sector. The photon kinematics 
thus remains unchanged and the decay of an initial photon into 
any number of final photons is only possible if all final
photons are collinear with the initial photon. 

The main part of Ref.~\cite{KoPi} is devoted to the
calculation of the relevant fermionic one-loop diagrams, to first order 
in the Lorentz-violating terms. The authors still find a vanishing 
amplitude for the splitting of an initial photon into two final ones. 
But they do find a non-zero amplitude for splitting into \emph{three} 
final photons, thereby establishing a difference with the case
of conventional quantum electrodynamics. 

The authors of Ref.~\cite{KoPi} close 
with some remarks on the possible physical 
significance of their result, i.e., whether or not the 
finite nonzero amplitude for photon triple splitting could lead to
observable effects. The obvious answer would appear to be negative, as
the phase space volume is zero for decay into three or more photons; 
cf. Refs.~\cite{Ha,Ad,WiSm}. 
But the authors argue in favor of a possible physical effect by
advocating an analogy to photon splitting via collinear parametric
down-conversion in optically active crystals.
Specifically, they quote an experiment reported in Ref.~\cite{GiMi}. 

The analogy is, however, misleading, for the following two reasons.
First, the experiment described in Ref.~\cite{GiMi} is based on 
phenomena induced by \emph{quadratically} nonlinear optics, as 
follows from their expressions for the nonlinear
polarization. Microscopically, this corresponds to the decay of
one initial photon with energy $\hbar\,\omega_\mathrm{\, p}$ (``pump'') 
into \emph{two} final photons with energies $\hbar\,\omega_\mathrm{\, s}$ 
(``signal'') and  $\hbar\,\omega_\mathrm{\, i}$ (``idler''), where 
$\omega_\mathrm{\, p} =\omega_\mathrm{\, s} +\omega_\mathrm{\, i}$. 
The decay into two photons has, of course, a non-zero phase space 
volume \cite{Ha,Ad,WiSm}.

Second, nonlinear optical phenomena like parametric down-conversion, 
parametric fluorescence, etc., almost always involve  
either the \emph{spatially inhomogeneous structure} or 
the \emph{nonstandard dispersion law}  
of the optical device used; cf. Ref.~\cite{Yar}.
It may, for instance, happen that the photon momentum conservation condition   
$\vec{p}_\mathrm{\, p} =\vec{p}_\mathrm{\, s} +\vec{p}_\mathrm{\, i}$ 
does not hold exactly (momentum being absorbed by the crystal)
or that the momenta of the initial and final photons are not perfectly collinear 
(consistent with the nonstandard dispersion law). 
These effects are, of course, absent for photons which propagate in a homogeneous 
vacuum and which have the standard vacuum dispersion law 
$p_0 = \omega\left(\vec{p}\:\right) = |\vec{p}\,|$,   
as is the case for the model of Ref.~\cite{KoPi}.        

A better physical analogy may be 
photon splitting in a constant external magnetic field \cite{Ad}. 
As long as this background
magnetic field is constant (i.e., time independent and homogeneous) and
the photons obey the standard vacuum dispersion law, 
the only possibility is the splitting of a single photon into two. 
The reason is that energy and momentum cannot be extracted from the 
constant background field and that 
the phase space volume for higher splittings is zero \cite{Ha,Ad,WiSm}.
Since the vacuum of the model studied in Ref.~\cite{KoPi} is perfectly
homogeneous (although no longer isotropic)
and the photons obey the standard dispersion law, the same reasoning
applies also in this case. [Possible quantum gravity effects which 
induce inhomogeneities (e.g., a spacetime foam) are not considered.]

We conclude that the comparison to parametric down-conversion as advocated by 
Ref.~\cite{KoPi} is inappropriate and that the probability for photon triple 
splitting is strictly zero, at least to quadratic order in the coefficients
of the Lorentz-violating terms considered. Indeed, there appears to be a 
kind of ``conspiracy'' in quantum electrodynamics models with
only the fermion sector modified, which keeps the photon stable by
setting either the amplitude or the phase space volume equal to zero.
For on-shell photon triple splitting, it may very well be necessary 
to have a modified photon sector; cf. Ref.~\cite{AdKl}. 
\vspace{0.5\baselineskip}\newline\noindent
\emph{Acknowledgements:} CA acknowledges support from the Austrian
START award project FWF-Y-137-TEC of N.J. Mauser.

\end{document}